\documentclass[prb,twocolumn,aps, superscriptaddress,floatfix,showpacs]{revtex4}

\usepackage{graphicx, epsfig}
\usepackage{amssymb,amsmath}

\def \hf{\tfrac{1}{2}}    

\def \ord{\mathcal{O}}

\def\lba{\left(}    \def\rba{\right)}

\newcommand{\bra}[1]{\langle\left.{#1}\right|}
\newcommand{\ket}[1]{\left|{#1}\right.\rangle}
\DeclareMathOperator{\tr}{tr}

\begin{document}

\title{Bipartite entanglement entropy in fractional quantum Hall states}

\author{O.S.~Zozulya}
\affiliation{Institute for Theoretical Physics, University of
Amsterdam, Valckenierstraat 65, 1018 XE Amsterdam, the Netherlands}

\author{M.~Haque}
\affiliation{Max-Planck Institute for the Physics of Complex
Systems, N\"othnitzerstr.~38, 01187 Dresden, Germany}

\author{K.~Schoutens}

\affiliation{Institute for Theoretical Physics, University of
Amsterdam, Valckenierstraat 65, 1018 XE Amsterdam, the Netherlands}

\author{E.~H.~Rezayi}
\affiliation{Dept. of Physics, California State University,
Los Angeles, California 90032, USA}

\date{May 29, 2007}

\begin{abstract}

We present a detailed analysis of bipartite entanglement entropies in
fractional quantum Hall (FQH) states, considering both abelian
(Laughlin) and non-abelian (Moore-Read) states. We derive upper bounds
for the entanglement between two subsets of the particles making up the
state. We also consider the entanglement between spatial regions
supporting a FQH state. Using the latter, we show how the so-called
topological entanglement entropy $\gamma$ of a FQH state can be extracted
from wavefunctions for a limited number of particles.

\end{abstract}


%03.67.Mn   Entanglement production, characterization, and manipulation
%71.10.-w   Theories and models of many-electron systems
%73.43.-f   Quantum Hall effects

\maketitle

\section{Introduction}

The fractional quantum Hall (FQH) states have long fascinated the
condensed-matter community due to their remarkable transport properties and
the exotic nature of their quasiparticle excitations. It is in the context of
FQH states that the notion of topological order in gapped two-dimensional
states first arose. \cite{WenNiu_PRB} Recently there has been enhanced
interest in FQH states with \emph{non-abelian statistics},
\cite{Moore-Read_NP91, Read-Rezayi_PRB99, Ardonne_Schoutens99} due to the
possibility of implementing quantum computation schemes topologically
protected from decoherence.  \cite{topological-quantum-computing} The unusual
features of FQH states have been notoriously difficult to characterize using
traditional condensed-matter concepts such as local order parameters and
$n$-point correlation functions.

In a separate development, recent years have seen growing understanding that
entanglement measures borrowed from the discipline of quantum information can
be useful in probing global features of quantum many-particle states.
\cite{VidalLatorreRicoKitaev_PRL03, Entng-in-spin-chains, Cardy_JStatMech04, Eisert_2002}
It is thus natural to ask what features of FQH states can be characterized by
entanglement measures.  In a recent short report,\cite{our-prl-07} three of
the present authors have shown that one such entanglement measure, the
\emph{bipartite entanglement entropy}, indeed elucidates the subtle
correlations and topological order in the simplest FQH states, those in the
so-called Laughlin sequence.  The bipartite entanglement entropy is defined by
partitioning the system under question into two blocks $A$ and $B$, and using
the reduced density matrix of one part (e.g., $\rho_A = \tr_B\rho$ obtained by
tracing over $B$ degrees of freedom) to calculate the von Neumann entropy $S_A
= -\tr[\rho_A\ln\rho_A]$.

In Ref.~\onlinecite{our-prl-07}, numerical calculations of the
entanglement entropy between two spatial regions allowed us to extract
from the Laughlin wavefunctions the so-called \emph{topological
entanglement entropy}, a concept introduced in
Ref.~\onlinecite{Preskill-Kitaev_PRL06, Levin-Wen_PRL06}.
[For brevity we write \emph{topological entropy} where no confusion
can arise.] In addition, we provided results on the entanglement
entropy between subsets of the particles making up the state. We showed
that such \emph{particle entanglements} are bounded by expressions
that manifest the exclusion statistics in the Laughlin states.

In this article, we present a systematic discussion of bipartite entanglement
entropies for FQH states, elaborating on our results in
Ref.~\onlinecite{our-prl-07}. In addition to the abelian Laughlin (L) states,
we consider a series of non-abelian FQH states: the Moore-Read (MR) (or
pfaffian) states. \cite{Moore-Read_NP91, Read-Rezayi96, Read_2000-overview} In
planar geometry, the respective wavefunctions are given by
\begin{eqnarray*}
& \Psi_{\rm L}(\{z_i\}) & =\prod_{i<j} (z_i-z_j)^m e^{- \sum_i |z_i|^2/4}
\nonumber \\
& \Psi_{\rm MR}(\{z_i\}) &= {\rm Pf} \left( \frac{1}{z_i-z_j} \right)
                    \prod_{i<j} (z_i-z_j)^m e^{- \sum_i |z_i|^2/4} \ ,
\end{eqnarray*}
with ${\rm Pf}$ denoting the antisymmetric Pfaffian symbol.
We shall here consider these same states in spherical geometry, so as
to eliminate boundary effects.

For both series of states, we derive upper bounds
$S_A^{\rm bound}$ for particle entanglement entropies. A marked difference
between the $m=3$ Laughlin state and the $m=2$ Moore-Read states is that
in the latter the leading correlations have a 3-body nature, whereas those
in the Laughlin states are 2-body effects. This difference is nicely
manifested in the leading terms of a $1/N$ expansion of the upper bounds
$S_A^{\rm bound}$, which are given by (see section III below for details)
\begin{eqnarray*}
&& m=3\ {\rm Laughlin}\ {\rm state:}
\nonumber\\
&& \qquad S_A^{\rm bound} = S^F_A - \frac{2}{3N} n_A(n_A-1) + \ldots
\nonumber\\[2mm]
&& m=2\ {\rm Moore-Read}\ {\rm state:}
\nonumber \\
&& \qquad S_A^{\rm bound} = S^F_A - \frac{3}{4N^2} n_A(n_A-1)(n_A-2) + \ldots
\end{eqnarray*}

Another marked difference between the abelian and non-abelian states
is in the value for the topological entropy $\gamma$. Comparing
the Laughlin and Moore-Read states at the same filling fraction $\nu=1/m$
we have
$$
\gamma_{\rm L} = \ln \sqrt{m} \ , \qquad
\gamma_{\rm MR} = \ln \sqrt{4m} \ ,
$$
the difference being due to the non-abelian nature of the Moore-Read
states (see section IV for details). In this article we extract values
for $\gamma$ directly from wavefunctions for a limited number of particles
in spherical geometry (up to $N=10$ for the $m=3$ Laughlin state and
up to $N=18$ for the $m=2$ Moore-Read state), finding values that are
consistent with the expected result. These results illustrate
how the entanglement entropy can be used in diagnosing the topological
order for a FQH state that is only known in the form of wavefunctions
for a limited number of particles, as is often the case in numerical
studies.

In Section \ref{sec_partition-choices} we give a general discussion
of possible bipartite entanglement measures in itinerant many body
systems, carefully distinguishing between \emph{particle} and
\emph{spatial} partitioning schemes.
In Section \ref{sec_pcle-partition} we present analytical and numerical
results for particle entanglement in the Laughlin and Moore-Read FQH
states. In this, the eigenvalue distribution of the reduced density
matrix $\rho_{n_A}$ plays a central role. In subsection
\ref{sec_pcle-RDM-n-corrln-fn} we relate the eigenvalue distribution for
$\rho_{n_A=2}$ to the two-particle correlation function $g_2(r)$.
In Section \ref{sec_spatial-n-gamma} we discuss spatial partitioning,
paying particular attention to the numerical procedure followed in
extracting the topological entropy $\gamma$.

While most results in this article are for fermionic FQH states
(meaning $m$ odd in the Laughlin sequence and $m$ even for the Moore-Read
states), we briefly comment on bosonic states in subsection
\ref{sec_bosonic}.

\section{Partition choices on the sphere}
\label{sec_partition-choices}

The entanglement entropy, being a \emph{bipartite} measure of
entanglement, depends on the particular partitions being
considered. Obviously, a many-particle system can be partitioned
in many ways.
Rather than asking which partition is the ``correct'' one, we find
it more useful to ask what information one can extract from
various kinds of partitioning.  Accordingly, we have partitioned
both the spatial degrees of freedom, and the particles themselves.
We find that both schemes are useful, for revealing distinct
features of the many-particle state.

The kinds of partitioning one is able to study depend on the
available degrees of freedom.  Our calculations are all performed
for FQH states in a spherical geometry. \cite{Haldane_FQHE_PRL83,
ArovasAuerbachHaldane_PRL88}  In this representation the fermions
are placed on a sphere containing a magnetic monopole.  The
magnetic orbitals of the relevant Landau level are then
represented as angular momentum orbitals; the total angular
momentum is half the number of flux quanta, $L={\hf}N_{\phi}$.
The $N_{\phi}+1$ orbitals are labeled either $l=0$ to $N_{\phi}$
or $L_z=-L$ to $+L$.  For $N$ particles at fractional filling
$\nu=1/m$, one finds the interesting FQH states for
$N_{\phi}=mN-S$, where $S$ is a finite-size shift.  The Laughlin
states appear at $S=m$ while for the Moore-Read states $S=m+1$.
The ``filling'' acquires the usual meaning $\nu =N/N_{\phi}$ only
in the thermodynamic limit.
The orbitals are each localized around a ``circle of latitude'' on
the sphere, with the $l=0$ orbital localized near one ``pole.''

Since the FQH wavefunctions on a sphere are obtained in terms of
orbital occupancies, one can either partition orbitals or
partition particles. Because of the spatial arrangement of the
orbitals, partitioning orbitals is in fact equivalent to
partitioning spatial regions.  The difference between spatial and
particle partitioning has not been stressed in the literature
because the most common systems studied (in the context of
entanglement entropies in many-particle states) are spin models,
for which there is no such distinction.  In the cases of itinerant
particles where there is a difference, the common default scheme
has been spatial partitioning.  In particular, conformal field
theory results on entanglement scaling \cite{HolzheyWilczek_NPB95,
Cardy_JStatMech04} and the distinction between gapless and gapped
states observed in entanglement scaling
\cite{VidalLatorreRicoKitaev_PRL03} actually pertain to the
blocking of \emph{space} rather than the particles or spins
themselves. The definition of the topological entropy for
two-dimensional topologically ordered states is also based on the
entanglement entropy between spatial blocks.
\cite{Preskill-Kitaev_PRL06,Levin-Wen_PRL06}

In previous work, \cite{our-prl-07} three of the present authors studied the
entanglement entropy between subsets of particles making up a Laughlin FQH
state.  We presented upper bounds and gave an interpretation in terms of
exclusion statistics. In this paper we extend these results to the Moore-Read
states, where the exclusion effects are more intricate.  We refer to
Refs.~\onlinecite{Tsinghua_FQHE_PRA02, EntngQHE_Lattore,
Pcle_entanglement_Santachiara} for other studies of
particle entanglement properties.

For orbital or spatial partitioning, we define block $A$ to be the
first $l_A$ orbitals, extending spatially from one pole of the
sphere out to some latitude.
In the thermodynamic limit, this is equivalent to choosing a
disk-shaped block $A$ within an infinite planar system.  In this
limit, since each orbital $l$ is associated with a wavefunction of
the form $z^le^{-|z|^2/4}$ in usual complex coordinate language, a
disk with $l_A$ orbitals has radius $\propto\sqrt{l_A}$.

The spatial arrangement of the orbitals constrains us to either
disc or ring-shaped spatial regions as the $A$ partition. Since
the orbital indexes do not give us access to the full
two-dimensional degrees of freedom, it is not possible to
experiment with the various kinds of topologically nontrivial
partitions suggested by Preskill and Kitaev,
\cite{Preskill-Kitaev_PRL06} Levin and Wen,
\cite{Levin-Wen_PRL06} and Furukawa and Misguich.
\cite{Misguich_dimer-gamma_dec06}  However, as we have reported
previously, \cite{our-prl-07} the spherical geometry is sufficient
to probe the topological entropy of FQH wavefunctions.

Comparing particle and spatial entanglement entropies, we remark
that the effect of correlations is opposite between the two cases.
For particle partitioning, the maximal entropy is realized for
uncorrelated fermions; correlations tend to lower $S_A$ from
the fermion bound $S_A^F$. Spatial entanglement, on the other hand,
is entirely due to correlations. The point is illustrated by
considering the $m=1$ Laughlin state, where the fermions are
uncorrelated. Here the particle entanglement entropy equals $S_A^F$,
while the spatial entanglement entropy vanishes.

\section{Entanglement for particle partitioning}
\label{sec_pcle-partition}

In this section we provide close upper bounds to the entropy of entanglement
between $n_A$ particles of the state and the remaining $n_B=N-n_A$ particles.
We also discuss the $n_A$-particle reduced density matrices
$\rho_{n_A}$ that arise in this context.

\subsection{Multiplet structure and particle entropy bounds }

For FQH states on a sphere, the $n_A$-particle reduced density matrices
$\rho_{n_A}$ commute with the total angular momentum operators
${\bf L}^2_{n_A}$ and $L^z_{n_A}$ of the selected $n_A$ particles.
This implies that the eigenvalues of $\rho_{n_A}$ are
organized in a multiplet structure of the corresponding $SU(2)$ algebra:
an eigenvalue for total angular momentum $L_{n_A}$ will be
$( 2 L_{n_A} + 1 )$-fold degenerate.

For $n_A=2$ fermions, each having angular momentum $L$, the 2-particle
states have total angular momenta $L_2=2L-1$, $2L-3$, $\ldots$, $1\, (0)$,
for $L$ integer (half-integer), giving a total number of $(2L+1)(2L)/2$
states. A naive upper bound to the entanglement entropy is thus
\begin{equation}
S_{n_A=2} \leq \ln \left[ (2L+1)(2L)/2 \right] .
\end{equation}
Inspecting the explicit structure of the fermionic Laughlin states
with $m=3,5,\ldots$, one finds that the eigenvalues corresponding to
2-particle states with $L_2=2L-1$, $2L-3$, $\ldots$, $2L-(m-2)$ all
vanish. The reason is that the correlations in the Laughlin states are
such that particles cannot come too close together. For example, if a
first fermion occupies the $l=0$ orbital, localized near the north pole,
the Laughlin wavefunction has zero amplitude for finding a second
fermion in orbitals $l=1$, $l=2$, $\ldots$, $l=m-1$. The highest
possible value of the angular momentum of the two fermions combined is
thus $L_2=L+(L-m)$. The remaining number of non-zero eigenvalues is
$(2L+(2-m))(2L+(1-m))/2$, leading to an improved bound on the entropy
$S_{n_A=2}$
\begin{equation}
S_{n_A=2} \leq \ln \left[ (2L+(2-m))(2L+(1-m))/2 \right]
\end{equation}
with $2 L=m(N-1)$ as before.
For $n_A>2$, the multiplet structures are more complicated and
we need to resort to a different method for finding a non-trivial
upper bound to the particle entropy. In the next subsection we give
a general derivation for both the Laughlin and the Moore-Read
series of fermionic FQH states.

\subsection{Upper bounds for fermionic states}

For $N$ fermionic particles, $n_A$ particles in the $A$ block, and the total
number of orbitals given by $N_\phi+1 = 2L+1$, fermionic statistics lead to an
obvious upper limit $S^{\rm F}_A$ to the entropy $S_A$
\begin{equation}
S_A \leq S^{\rm F}_A = \ln \left( \begin{array}{c} N_\phi+1 \\ n_A
\end{array} \right) \ .
\end{equation}
In the FQH states the correlations are such that the particles
avoid each other and the entropy is further reduced. To obtain a handle
on this, one may reason as follows. The model FQH states
in the Laughlin and Moore-Read series can be characterized as zero-energy
eigenstates of a Hamiltonian penalizing pairs and/or triplets of particles
coming to the same position. After tracing out the coordinates for the $B$ set,
the dependence on those in the $A$ set is such that one still has a
zero-energy eigenstate. However, the number of orbitals
available to the $A$ particles is larger than what is needed to make the
model FQH state in the $A$ sector, and one instead has a certain number
of quasi-holes on top of the $A$ set model state. The total ground state
degeneracy for this situation has been studied in the literature:
see Ref.~\onlinecite{Read-Rezayi96} for the Laughlin and Moore-Read states
and Ref.~\onlinecite{NAcounting} for the Read-Rezayi and Ardonne-Schoutens
series of non-abelian FQH states.

For the Laughlin states the details are as follows. The $N$-particle
Laughlin state is realized on a total of $N_\phi+1$ Landau orbitals,
corresponding to $N_\phi=m(N-1)$ flux quanta. The Laughlin state for $n_A$
particles would need $N_\phi^A=m(n_A-1)$ flux quanta; we thus have
an excess flux of $\Delta N_\phi= N_\phi-N_\phi^A=m(N-n_A)$. With the
Laughlin gauge argument this corresponds to the presence of
$n_{\rm qh}= \Delta N_\phi$ quasi-holes over the groundstate.
According to Ref.~\onlinecite{Read-Rezayi96} each of the quasi-holes has
a number of $n_A+1$ effective orbitals to choose from, with bosonic
counting rules (meaning that two or more quasi-holes can be in the same
effective orbital). This gives a number of quasi-hole states equal to
$$
\left( \begin{array}{c} (n_A+1) + n_{\rm qh} - 1
                      \\ n_{\rm qh}
       \end{array} \right) \ ,
$$
leading to the following upper bound to the entropy $S_A$
\begin{equation}
S^{\rm bound}_A = \ln
\left( \begin{array}{c}
N_\phi+1-(m-1)(n_A-1) \\ n_A
\end{array} \right) \ .
\label{LaughlinBound}
\end{equation}
We remark that this expression has a clear interpretation in
terms of exclusion statistics: the counting factor in
Eq.~(\ref{LaughlinBound}) gives the number of ways $n_A$ particles
can be placed in $N_\phi+1$ orbitals, in such a way that a particle
placed in a given orbital $l$ excludes particles from orbitals $l'$
with $|l-l'|<m$.

In a $1/N$ expansion we find (assuming $n_A\ll N$)
\begin{eqnarray}
\lefteqn{S^{\rm F}_A - S^{\rm bound}_A=}
\nonumber \\[2mm]
&& \frac{1}{N} \frac{m-1}{m} n_A (n_A-1)
\nonumber \\[2mm]
&& + \frac{1}{N^2} \frac{m-1}{2 m^2} n_A (n_A-1)
     [2m + (n_A-1)(m+n_A-4)]
\nonumber \\[2mm]
&& + \mathcal{O}(1/N^3)
\label{Laughlin1overN}
\end{eqnarray}

The particle entropy reaches a maximum for $n_A=N/2$.
For this case our Eq.~(\ref{LaughlinBound}) gives, in
the limit of large $N$,
\begin{equation}
S_{n_A=N/2} \leq N [(m+1) \ln(m+1)- m \ln(m)]/2 \ .
\end{equation}
This bound is sharper than a bound recently presented
in Ref.~\onlinecite{EntngQHE_Lattore}, which gives a larger coefficient
for the linear-in-$N$ behavior.

For the fermionic Moore-Read states at $\nu=1/m$,
with $m=2,4,\ldots$, we can reason in a similar way,
with now $N_\phi=m(N-1)-1$.
As for the Laughlin states we have an excess flux of
$\Delta N_\phi = N_\phi-N_\phi^A=m(N-n_A)$
but now the number of quasi-holes is twice this number due to
the fact that the fundamental quasi-holes correspond to half
a flux quantum. Thus, $n_{\rm qh}=2\Delta N_\phi$. We now take
from Ref.~\onlinecite{Read-Rezayi96} the following result for
the total quasi-hole degeneracy
\begin{equation}
 \sum_{F \equiv n_A \mod\ 2}^{n_A}
     \left( \begin{array}{c} n_{\rm qh}/2 \\ F \end{array} \right)
     \left( \begin{array}{c} (n_A -F)/2+ n_{\rm qh}
                           \\ n_{\rm qh} \end{array} \right) \ .
\label{MRBound}
\end{equation}
This gives us an upper bound $S^{\rm bound}_A$ as before.

Putting $m=2$, one easily checks that $S^{\rm bound}_A$
coincides with $S_A^{\rm F}$ for $n_A=2$. In a $1/N$ expansion,
the leading deviation from $S_A^F$ is a 3-body term at order $1/N^2$,
\begin{equation}
S^{\rm F}_A - S^{\rm bound}_A =
\frac{1}{N^2} \frac{3}{4} n_A (n_A-1)(n_A-2) + \ldots
\label{MRm=21overN}
\end{equation}
This result nicely illustrates the fact that the leading correlations
in the $m=2$ Moore-Read state have a 3-body character: the wave-function
vanishes if at least three particles come to the same position.

For $m\neq 2$ the leading correlations do have a 2-body character, as
for the Laughlin states,
\begin{equation}
S^{\rm F}_A - S^{\rm bound}_A =
\frac{1}{N} \frac{m-2}{m} n_A (n_A-1) + \ldots
\label{MR1overN}
\end{equation}

Inspecting the particle entanglement at $n_A=N/2$ and for $N$ large,
our bound implies that for the $m=2$ Moore-Read state
\begin{equation}
S_{n_A=N/2} \leq 1.044 N \ .
\end{equation}
This bound is reduced from the Fermi bound
$S^F_A \sim N (4\ln 4 -3 \ln 3)/2 \sim 1.125\, N$, but it is larger
than the bound for the $m=2$ (bosonic) Laughlin state, which has
asymptotic form $N (3\ln 3 -2 \ln 2)/2 \sim .955\, N$. This
indicates that, at equal filling $\nu=1/2$, the particles in a
Moore-Read state are more entangled than those in a Laughlin state.

The quasi-hole counting rules for the order-$k$ clustered
spin-polarized (Read-Rezayi) and spin-singlet (Ardonne-Schoutens)
states are all known in the literature. \cite{NAcounting}
They can be used to generalize the upper bounds on particle
entanglement entropy given in this subsection to these more
intricate non-abelian FQH states.

\subsection{Bosonic quantum Hall states} \label{sec_bosonic}

We briefly comment on the case of bosonic FQH states.
The realization that a rapidly rotating Bose gas may eventually
enter a regime of bosonic quantum Hall states motivates the
theoretical study of the effects of bosonic statistics.

We consider bosonic Laughlin states at filling fraction
$\nu=\frac{1}{m}$ with $m=2,4,\ldots$. The naive upper bound to
the the entropy associated to placing $n_A$ bosons in $N_\phi+1$
orbitals is
\begin{equation}
S^{\rm B}_A = \ln \left( \begin{array}{c}
    N_\phi + n_A\\ n_A \end{array}
                   \right)
\end{equation}
The expression for $S^{\rm bound}_A$ remains unchanged,
giving the following leading correction in a $1/N$ expansion
\begin{equation}
S^{\rm B}_A - S^{\rm bound}_A = \frac{1}{N} n_A (n_A-1) + \ldots
\label{bLaughlin1overN}
\end{equation}

For a bosonic Moore-Read state, with filling fraction
$\nu=1/m$ with $m=1,3,\ldots$, the leading $1/N$ correction
becomes
\begin{equation}
S^{\rm B}_A - S^{\rm bound}_A
= \frac{1}{N} \frac{m-1}{m} n_A (n_A-1) + \ldots
\label{bMR1overN}
\end{equation}
In the case $m=1$ the leading correlations have 3-body
character, leading to the vanishing of the leading $1/N$
correction.

\subsection{Numerical results}

In deriving the upper bound $S_A^{\rm bound}$ we relied on the
fact that a certain number of eigenvalues of the reduced density
matrix vanish. The bounds would be exact if all non-zero
eigenvalues were equal, but since they are not the bounds
overestimate the actual values for the entropies.

\begin{figure}
\centering
 \includegraphics*[width=0.95\columnwidth]{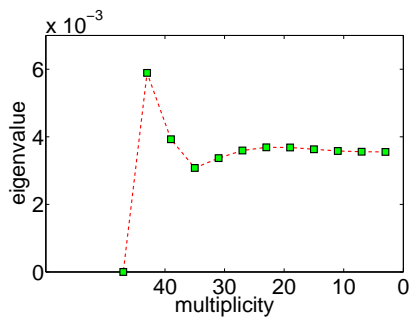}
\caption{  \label{fig_2-pcle_eigen}
(Color online)~
Eigenvalues for the 2-particle reduced density matrix, plotted against
their multiplicities, for $N=9$ particles in the $m=3$ Laughlin state.
}
\end{figure}

Fig.~\ref{fig_2-pcle_eigen} plots the eigenvalues for the $n_A=2$-particle
reduced density matrix for $N=9$ particles on a sphere in the $m=3$ Laughlin
state, for which the single particle angular momentum is $L=12$. The
horizontal axis represents the degeneracy $2 L_2 + 1$ of the eigenvalues, in
descending order. The eigenvalue at $L_2=2L-1=23$, with degeneracy 47,
vanishes; the non-zero eigenvalues show some scatter around an asymptotic
value. Due to this scatter the entropy $S=5.509$ is somewhat lower than the
upper bound $S_A^{\rm bound}= 5.533$.

\begin{figure}
\centering
 \includegraphics*[width=0.95\columnwidth]{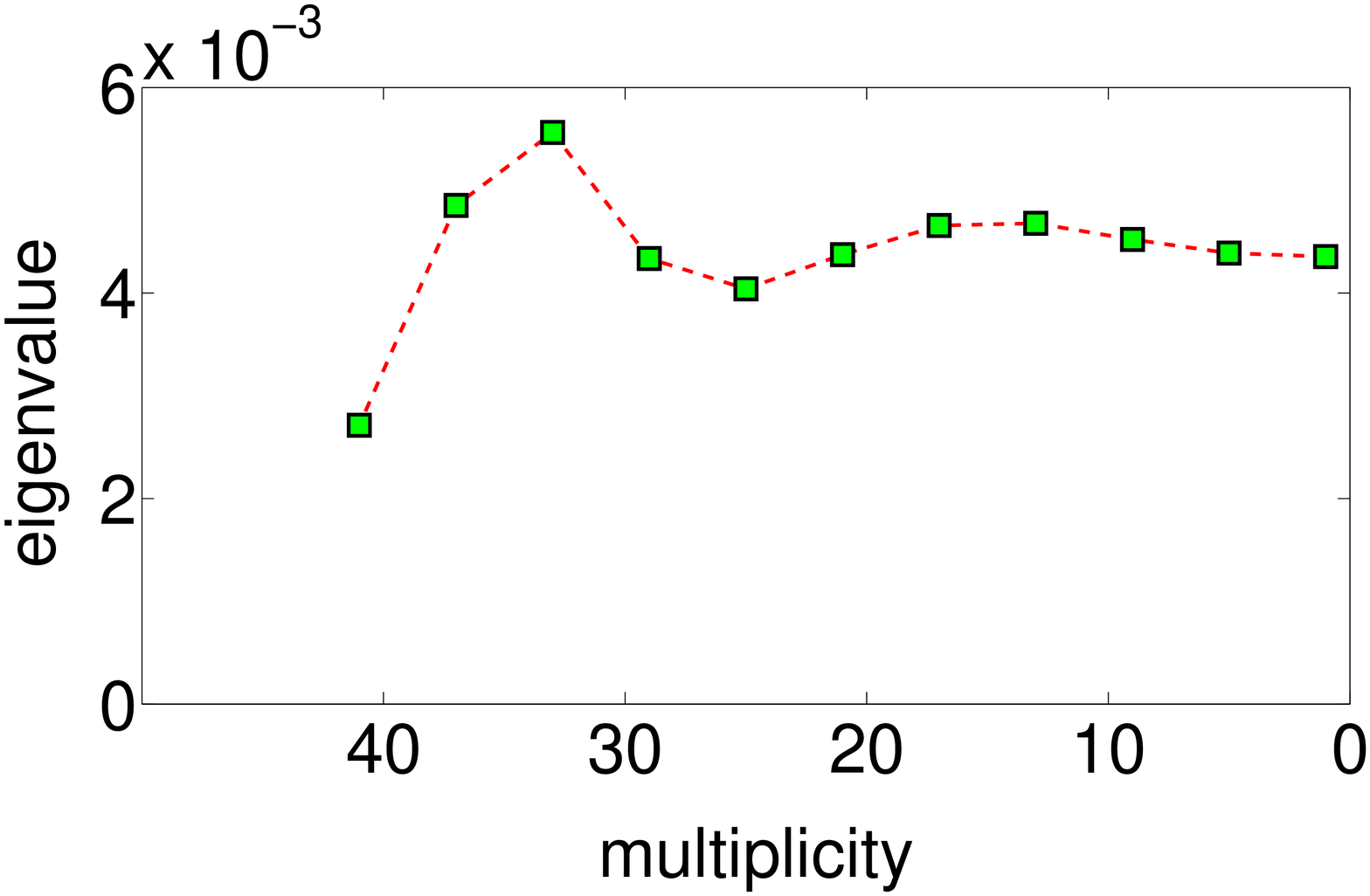}
\caption{  \label{fig_2-pcle_pf_eigen}
(Color online)~
Eigenvalues for the 2-particle reduced density matrix, plotted against
their multiplicities, for $N=12$ particles in the $m=2$ Moore-Read state.
}
\end{figure}

An important difference between the $m=3$ Laughlin and the $m=2$
Moore-Read states is the absence of vanishing eigenvalues for
the 2-particle reduced density matrix. The eigenvalue distribution
shown in Fig.\ \ref{fig_2-pcle_pf_eigen} illustrates this point.

In the $m=2$ Moore-Read state, there are vanishing eigenvalues
in the reduced density matrix of $n_A\geq 3$ particles. The number of
nonzero eigenvalues predicted by Eq.~(\ref{MRBound}) agrees with
numerical results. For example, for $n_A=3$ and $N=10$ particles there
are 770 nonvanishing eigenvalues, in agreement with Eq.~(\ref{MRBound}).

\begin{figure}
\centering
 \includegraphics*[width=0.95\columnwidth]{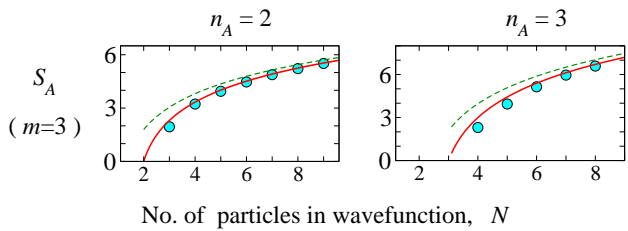}
\caption{  \label{fig_pcle_lghln_23}
(Color online)~
Entanglement entropy for $n_A=2$ and $n_A=3$ particles for the
$m=3$ Laughlin state. Dots are numerical exact values, the dotted
line represents $S_A^F$ and the solid curve is the bound
$S_A^{\rm bound}$. }
\end{figure}

\begin{figure}
\centering
 \includegraphics*[width=0.95\columnwidth]{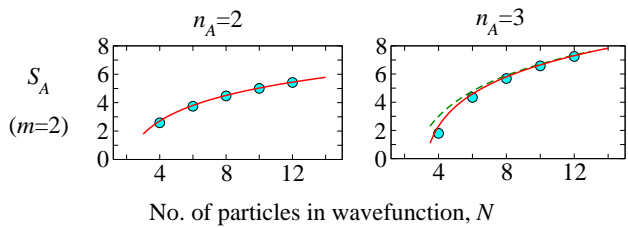}
\caption{  \label{fig_pcle_pf_23}
(Color online)~
Entanglement entropy for $n_A=2$ and $n_A=3$ particles for the
$m=2$ Moore-Read state. Dots are numerical exact values, the dotted
line represents $S_A^F$ and the solid curve is the bound
$S_A^{\rm bound}$.}
\end{figure}

In Figs.~\ref{fig_pcle_lghln_23}, \ref{fig_pcle_pf_23} we compare numerically
computed particle entanglement entropies with the bounds derived above.

\subsection{Corrections to $S^{\rm bound}_A$ due to eigenvalue spread}

It is interesting to consider in some detail the deviation between
the bounds $S_A^{\rm bound}$ and the actual entropies computed
numerically. As mentioned above, this deviation arises from
the fact that the non-zero eigenvalues of the reduced density matrices
are not all equal.

To estimate the effect on $S_{n_A}[N]$ of the spread in the non-zero
eigenvalues, we do a rough modeling of the eigenvalue distribution (Fig.\
\ref{fig_2-pcle_eigen}) of $\rho_{n_A=2}$ for the $m=3$ Laughlin state.  For
this case the number $D$ of non-zero eigenvalues is $D = (3N-4)(3N-5)/2$. If
these nonzero eigenvalues were all equal (to $1/D$), the entanglement entropy
would have the maximum value $S_D = \ln{D}$, which is the predicted upper
bound (\ref{LaughlinBound}). We now take into account the deviations from
$1/D$, guided by Fig.\ \ref{fig_2-pcle_eigen}, with the following toy
distribution: we take $D_0 $ out of $D$ of the eigenvalues to be equal to
$\alpha/D$, with $\alpha > 1$, while the rest of the eigenvalues are at value
$\beta/D$, $\beta < 1$, such that the sum of eigenvalues is unity. Assuming
$D_0/D \ll 1$ leads to
\begin{equation}
S \approx \ln D - (\alpha \ln \alpha - \alpha + 1) \frac{D_0}{D}  \ .
\end{equation}
Guided by the eigenvalue distributions in Fig.~\ref{fig_2-pcle_eigen}, we
assume that $D_0/D$ is of order $1/N$; for concreteness we put $D_0$ equal to
the multiplicity of the largest eigenvalue, which is $6N-7$. Taking $\alpha$
between 1.2 and 1.5 (as observed for the largest available Laughlin
wavefunctions) gives a $1/N$ correction in the entropy with coefficient in the
range 0.03 -- 0.14.

Fitting the difference $S_A^F-S_{n_A=2}$ to a form $a/N + b/N^2$ gives
a coefficient $a \simeq 1.38$. The vanishing eigenvalues account
for $a=4/3$, see Eq.~(\ref{Laughlin1overN}), and we see that the remaining
difference $\delta a \sim 0.05$ is consistent with the $1/N$ correction
due to the spread in the non-zero eigenvalues.

We made similar estimates for a the $m=2$ Moore-Read state with up
to $N=12$ particles, where the eigenvalues are all non-zero and
$S_A^{\rm bound}$ agrees with $S_A^F$. In this case the deviation
between data and bound show a $1/N$ dependence with a coefficient
of about 0.14.

These considerations are of some general interest, as they make the point that
a $1/N$ expansion of particle entanglement entropies are indicative of
correlations in a many-body state. In the concrete case studied here, the
sizeable value of the leading $1/N$ correction in the Laughlin state indicates
strong 2-body correlations, while the small value for the Moore-Read state
indicates the absence of such correlations.

\subsection{Reduced density matrices and correlation functions}
\label{sec_pcle-RDM-n-corrln-fn}

Since the $n_A$-particle reduced density matrices $\rho_{n_A}$ are obtained by
integrating out all but $n_A$ of the particles, one expects these matrices to
be related to the $n_A$-particle correlation functions.  In this subsection we
study this relation for the case $n_A=2$.  In particular, the eigenvalue
distributions of $\rho_{n_A=2}$ in
Figs.~\ref{fig_2-pcle_eigen},~\ref{fig_2-pcle_pf_eigen}, although discrete,
are reminiscent of the well-known two-particle correlation functions $g_2(r)$
for Laughlin and Moore-Read states. We will show that the eigenvalue
distributions are in fact very closely related to the correlation functions --
the eigenvalue distribution function is a kind of discretized version of
$g_2(r)$.

The two-particle correlation function $g_2(r)$ is conventionally
defined as
\begin{multline}
g_2(r) = \frac{N(N-1)}{n^2} \times \\
\frac{ \int d^2 r_3 \cdots d^2 r_n \Psi^* (0,r,r_3,\cdots, r_n)
  \Psi(0, r, r_3, \cdots, r_n)}{ \langle \Psi | \Psi \rangle } \: ,
\end{multline}
where $N$ is the number of particles and $n$ is a density, which is chosen
such that $g_2(r) \stackrel{r \rightarrow \infty}{=} 1$.

We express the 2-particle reduced density matrix $\rho_2$ on a sphere
in a basis of polar spherical coordinates. Because of the rotational
symmetry it should be a function of the angular distance $\theta$
between the two particles. In Appendix~\ref{eigenvalues_correlations}
we show that $\rho_2(\theta)$ can be written in the form
\begin{equation}
\rho_2(\theta) = \sum_{l} \lambda_l \frac{2l+1}{4\pi}
R_l(\theta) \; .
\label{rho2_theta}
\end{equation}
In this expression, $\lambda_l$ is the eigenvalue with multiplicity
$2l+1$, corresponding to the total angular momentum of the two particles
equal to $l$. The functions $R_l(\theta)$ are explicitly given
in Eq.~(\ref{Rtheta}).

\begin{figure}
\centering
\includegraphics*[width=0.95\columnwidth]{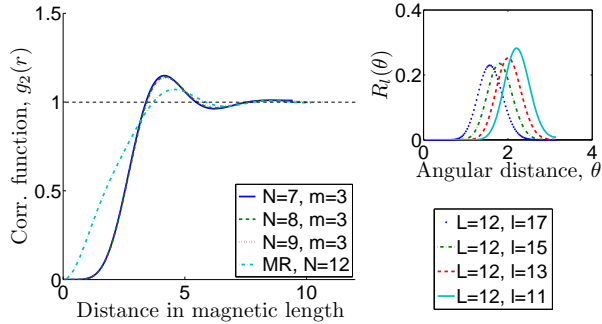}
\caption{  \label{fig_g2_Lghln_Pf}
(Color online)~
Two-particle correlation as a function of distance in units of the
magnetic length, for the $m=3$ Laughlin state with $N=7,8,9$ particles
and for the $m=2$ Moore-Read state with $N=12$.}
\end{figure}

Since the distance $r$ between two particles is simply equal to
$r = R\, \theta$, with $R$ the radius of the sphere, the 2-particle
correlation function $g_2(r)$ is directly proportional to $\rho_2(\theta)$.
Through Eq.~(\ref{rho2_theta}) it is expressed as a transfrom from $l$
space to $\theta$ space, with basis functions $R_l(\theta)$.
In Fig.~\ref{fig_g2_Lghln_Pf}, we show some curves for $\rho_2(\theta)$;
they agree with known results.

As illustrated in the inset to Fig.~\ref{fig_g2_Lghln_Pf}, the basis
functions $R_l(\theta)$ have a peak structure, with the position of
the peak depending on the total angular momentum $l$. Large values of
$l$ correspond to small angular distances and vice versa.
This is easy to understand from the following classical picture. When two
particles with angular momenta $L$ have total angular momentum $2L$, the
corresponding vectors $\vec{L}_1$ and $\vec{L}_2$ should point a the same
direction. The angular momentum vector points at the position of the particle
on a sphere, therefore two particle should be close to each other. On the
other hand, if total angular momentum is zero then the angular momentum
vectors should point into opposite directions. This means that particles
are placed at the opposite sides of the sphere.

The fact that the $R_l(\theta)$ are localized functions, peaked at $\theta$
values monotonically decreasing with $l$, indicates that $g_2(r)$ curve is
simply a continuous form of the $\lambda_l$ versus descending-$(2l+1)$ curves of
Figs.~\ref{fig_2-pcle_eigen} and \ref{fig_2-pcle_pf_eigen}.  The similarity
between the discrete $\lambda_l$ and the continuous $g_2(r)$ curves is not
accidental.

At small distances $R_l(\theta) \propto (\theta^2)^{2L-l}$.  For a Laughlin
state the lowest value of $2L-l$ is $m$, thus $\rho_2(\theta) \propto
\theta^{2m}$. This behavior is a direct consequence of the vanishing of
eigenvalues with the largest multiplicities.  For the $m=2$ Moore-Read state
the lowest value of $2L-l$ is $1$ because there are no vanishing
eigenvalues. Therefore at small distances $\rho_2(\theta) \propto \theta^{2}$.

Of course, our observations on the 2-particle correlations agree with
known results; our main point has been to stress the intimate relation
with the eigenvalue distribution of the 2-particle reduced density matrices.

\section{Spatial entanglement and topological entropy}
\label{sec_spatial-n-gamma}

We now turn to dividing the Landau level orbitals into two blocks and
calculating the entropy of entanglement between them.  In a previous
publication, \cite{our-prl-07} we used this scheme to extract the topological
entropy of the Laughlin state.  Here, we mainly focus on the ($m=2$)
Moore-Read state.  After reviewing the topological entropy $\gamma$ and the
total quantum dimension $\mathcal{D}$, especially in the context of the
Moore-Read state (\ref{subsec_gamma-review}), we detail some issues with
taking the thermodynamic limit necessary for extracting $\gamma$ from
numerical data (\ref{subsec_extrapoln-issues}), and then present our numerical
results (\ref{subsec_numerical-gamma}).  We also present observations on the
spectral structure of the reduced density matrices
(\ref{subsec_orbital-RDM-spectrum}).

\subsection{Topological entropy for the Moore-Read state}
\label{subsec_gamma-review}

For spatial partitioning of many-particle states, the general rule (``area
law'') is that the entanglement entropy scales as the size of the boundary
between the $A$ and $B$ blocks.  \cite{Srednicki_PRL93} Subtle information
about the nature of the many-particle state can be provided by the presence or
absence of logarithmic corrections, values of coefficients, or subleading
terms in this basic relationship.  For topologically ordered states in two
dimensions, the following theorem has been presented recently
\cite{Preskill-Kitaev_PRL06, Levin-Wen_PRL06} concerning the scaling of
entanglement entropy between spatial partitions.  If $L$ is the length of the
boundary between the two blocks, the entanglement entropy scales as $S_A =
\alpha{L} -\gamma +\ord(L^{-1})$.
As usual the scaling law applies to situations where $A$ is large
and the total system is infinite.
The subleading term $\gamma$ is called the \emph{topological entanglement
entropy}. The striking result of Refs.\ \onlinecite{Preskill-Kitaev_PRL06,
Levin-Wen_PRL06} has been that it can be expressed as the logarithm of a
quantity $\mathcal{D}$ known as the \emph{total quantum dimension} of the
topological field theory describing the topological order of the state. The
total quantum dimension is given by
\begin{equation}
\mathcal{D}=\sqrt{\sum_i d_i^2} \ ,
\end{equation}
where the $d_i$'s are the quantum dimensions of the individual
sectors making up the topological field theory. These quantum dimensions
are set by fusion rules of the fundamental anyons in the field theory.

The topological field theory for a $\nu=1/m$ Laughlin state has a
fundamental anyon (of fractional charge $q=e/m$), which generates $m$
abelian sectors. The quantum dimension $d_i$ is unity in all sectors
so that $\mathcal{D}=\sqrt{m}$ for the $\nu=1/m$ Laughlin state.
For $m=3$ this gives $\gamma=\ln\sqrt{3}\simeq0.55$.

For states with non-abelian quasiparticles, the situation is more interesting
because some anyon sectors contribute $d_i>1$. Details for some examples have
been provided in Refs.\ \onlinecite{Preskill-Kitaev_PRL06,
FendleyFisherNayak_JStatPhys07}.  In particular, for the $m=2$ Moore-Read
state, there are six sectors (two each of quasiparticles denoted by $I$,
$\sigma$, $\psi$) which contribute $d_I=1$, $d_{\sigma}=\sqrt{2}$,
$d_{\psi}=1$, leading to $\mathcal{D}=\sqrt{8}$ and $\gamma\simeq1.04$.  The
non-abelian nature shows up in the fact that $\gamma$ is larger than
$\ln\sqrt{6}$, six being the degeneracy of the $m=2$ Moore-Read state on the
torus.\\

A compact general expression for the total quantum dimension for a
Read-Rezayi state with order-$k$ clustering and at filling fraction
$\nu=k/(kM+2)$ is
(see also Ref.\ \onlinecite{FendleyFisherNayak_JStatPhys07})
\begin{equation}
\mathcal{D}_{\rm RR}[k,M]=
\frac{\sqrt{(k+2)(kM+2)}}{2 \sin(\pi /(k+2))} \ .
\label{scriptDRR}
\end{equation}
It includes the Laughlin states ($k=1$, $M=m-2$) and the Moore-Read
states ($k=2$, $M=m-1$) as special cases. For a general spin-singlet
non-abelian FQH state with order $k$ clustering and filling fraction
$\nu=2k/(2kM+3)$, the result is
\begin{equation}
\mathcal{D}_{\rm AS}[k,M]=
\frac{(k+3)\sqrt{(2kM+3)}}{16 \cos(\pi/(k+3)) \sin^3(\pi /(k+3))} \ ,
\label{scriptDAS}
\end{equation}
giving $\gamma\simeq1.62$ for the paired spin-singlet state
($k=2$, $M=1$) at $\nu=4/7$.

In general, the $M$-dependence of these expressions for total
quantum dimensions is linked to the ground state degeneracy in torus
geometry. Denoting the latter by $\#[k,M]$ we have the relation
\begin{equation}
\mathcal{D}[k,M]= \mathcal{D}[k,0] \sqrt{\frac{\#[k,M]}{\#[k,0]}} \ ,
\end{equation}
The conformal field theories underlying the states at $M=0$ are
of Wess-Zumino-Witten type ($SU(2)_k$ for the RR states and $SU(3)_k$
for the AS series) and the quantities $\mathcal{D}[k,0]$ can be
expressed in the modular $S$-matrix for these WZW models.

\subsection{$\gamma$ from sphere calculations: extrapolations}
\label{subsec_extrapoln-issues}

First of all, we note that, since our degrees of freedom are ordered
essentially one-dimensionally, we cannot use one of the two-dimensional
schemes proposed previously
\cite{Preskill-Kitaev_PRL06, Levin-Wen_PRL06, Misguich_dimer-gamma_dec06}
in which an appropriate
addition/subtraction of the entanglement entropies of several regions cancels
the boundary parts of the entropy ($S_A\rightarrow\alpha{L}-\gamma$) leaving
the subleading term $\gamma$.  With the orbital degrees of freedom on a
sphere, we can choose only regions corresponding to disks, concentric rings,
and combinations thereof.  Any combination of entropies of disk- and ring-like
regions that cancels out the boundary terms also unfortunately cancels out the
$\gamma$ term.

We are thus led to using directly the scaling law,
$S_A\xrightarrow{L\rightarrow\infty}\alpha{L}-\gamma$.  Our choice of block
$A$ as the first $l_A$ orbitals, extending spatially from one pole out to some
latitude, corresponds to a disk-shaped block only in the thermodynamic limit.
The block area is proportional to the square of $\sqrt{l_A}$ while its
boundary is proportional to $\sqrt{l_A(N_{\phi}+1-l_A)}$; these are equivalent
only in the same $N\rightarrow\infty$ limit.  One way to numerically access
the thermodynamic limit is to take the entanglement entropy of $l_A$ orbitals
with the rest, for accessible wavefunctions of various sizes $N$, and then
take the $N\rightarrow\infty$ limit.  The $S_{l_A}(N\rightarrow\infty)$ versus
$\sqrt{l_A}$ points thus obtained should then follow a linear curve at large
$l_A$, whose vertical intercept gives the topological entropy.  Results
following this procedure were provided for the $\nu=1/3$ Laughlin state in our
earlier paper; \cite{our-prl-07} here we will focus on the Moore-Read state.
The extrapolation of $S_{l_A}(N)$ values to the thermodynamic limit is a
tricky issue. We therefore discuss the extrapolation in some detail here,
providing some general results.

We are interested in the function $S_{l_A}(x)$, where $x=1/N$.  We have access
to $S_{l_A}(x_i)$ at several integer values of $N$, and would like to estimate
$S_{l_A}(0)$.  For each dataset that we have access to (each $l_A$; both
Laughlin and Moore-Read), we note the following: the $S_{l_A}(x_i)$ versus
$x_i$ values form a \emph{monotonic} curve and this curve gets flatter (slope
magnitude decreases) with decreasing $x$.  Two examples can be seen in the
inset to Fig.~\ref{fig_extrapolatn}.  In other words, the first and second
derivatives of the $S_{l_A}(x)$ function have the same sign and neither
derivative changes sign.

Motivated by the above observations, we provide the following result.
Assuming only that the signs of the first two derivatives of the $S_{l_A}(x)$
function are the same and that the signs remain unchanged until $x=0$, we
have:
\begin{enumerate}
\item The value $S_0 = S_{l_A}(x_0)$ corresponding to the smallest value $x_0$
  of the available $x_i$ is a strict lower (upper) bound for $S_{l_A}(0)$ if
  the $S_{l_A}'(x)$ is negative (positive).
\item The intercept found by connecting the $S_{l_A}(x)$ corresponding to the
  smallest two $x_i$ values ($x_0$, $x_1$), namely
\[
S_1 = S_{l_A}(x_0)\lba 1-\frac{x_0}{x_1-x_0}\rba
+S_{l_A}(x_1)\lba\frac{x_0}{x_1-x_0}\rba \; ,
\]
is a strict upper (lower) bound if  $S_{l_A}'(x)$ is negative (positive) and
$S_{l_A}''(x)$ is positive (negative).
\end{enumerate}
The limits $S_0$ and $S_1$ thus obtained give us conservative bounds for the
required entanglement entropies in the thermodynamic limit,
$S_{l_A}(N\rightarrow\infty)$.  To obtain a sharper extrapolation, one can use
various polynomial extrapolations and take the average, as done in our earlier
work. \cite{our-prl-07} Here, we improve the extrapolation by using the
extrapolation algorithm of Bulirsh and Stoer (BST algorithm), based on
rational polynomial fraction approximations. \cite{xtrap_BulirshStoer64,
xtrap_HenkelSchuetz88}

\begin{table}
\begin{ruledtabular}
\begin{tabular}{llll}
$\omega =$ 2.0  & & & \\
\hline
3.46601 & 5.21183 & 4.69023  & 4.77647  \\
3.80408 & 4.92355 & 4.74756  &          \\
4.01821 & 4.84476 &          &          \\
4.16743 &         &          &        \\
\hline\hline
$\omega =$ 3.0002  & & & \\
\hline
3.46601  &  4.56058 & 4.51976  & 4.55722  \\
3.80408  &  4.53643 & 4.59865  &          \\
4.01821  &  4.56761 &          &          \\
4.16743  &          &          &        \\
\hline\hline
$\omega =$ 2.5926  & & & \\
\hline
3.46601  & 4.74638 & 4.58027  & 4.65108  \\
3.80408  & 4.65109 & 4.65108  &          \\
4.01821  & 4.65108 &          &          \\
4.16743  &         &          &        \\
\hline
\end{tabular}
\end{ruledtabular}
\caption{\label{table_BST-1}
Extrapolation using the BST algorithm, using three different $\omega$ values
  for the same initial dataset (first column), corresponding to
  $S_{l_A=11}(N)$ values for $N=$ 12,14,16, 18.  In the last case, $\omega$
  has been tuned to ensure that the sets obtained after first and second
  iterations converge to the same value, i.e., the lowest elements of the
  second and third columns are the same. }
\end{table}

The BST procedure involves successive transformations of the original dataset,
leading to transformed datasets successively smaller by one element.  If the
free parameter ($\omega$) of the algorithm is chosen properly, the successive
sets will be more and more convergent and will eventually converge to the
$x=0$ value.  The parameter $\omega$ is chosen to optimize convergence;
how precisely this is done is a non-trivial issue, especially for
large datasets.

In the present case where the initial dataset consists of bipartite
entanglement entropies, it is particularly desirable to be able to use a small
number of initial $S_{l_A}(N)$ values, corresponding to the largest available
$N$.  This is because the symmetry $S_{l_A}=S_{N_{\phi}+1-l_A}$ makes the
$S_{l_A}(N)$ data meaningless for $N\rightarrow\infty$ extrapolation when
$l_A$ gets close to $N_{\phi}/2 = \hf(mN-S)$.  In other words, the
$S_{l_A}(N)$ values used should not be too close to the peak at the midpoint
of the respective $S_{l_A}(N)$ versus $l_A$ curve (e.g., Figure 1 of
Ref.~\onlinecite{our-prl-07}).  This restricts us to the largest few $N$
values.

In Table \ref{table_BST-1}, we demonstrate some possible choices of $\omega$
starting from four $S_{l_A}(N)$ values.  The top example uses $\omega=2$,
which is a common choice used for extrapolation-based numerical integration
algorithms.  In the second example, $\omega$ is tuned to give the ``most
converged'' set after the first iteration, by choosing $\omega$ to minimize
the standard deviation of the second column.  We find that this procedure does
not always lead to great stability for the subsequent iteration results.

In the last example, we have tuned $\omega$ to give the same last value for
the sequences obtained after the first and the second iterations.  This also
guarantees the third (final) iteration to give the same value (Table
\ref{table_BST-1}), which indicates that the convergence is very good.  This
procedure has the added advantage that actually only the last three points of
the
initial dataset are used to determine the values tuned to be equal.  We thus
get a nicely converged estimate based only on three initial points, which is
highly desirable as explained above.  This is therefore our method of choice
for tuning the free parameter $\omega$ in applying the BST algorithm for our
problem.

As a side note, we remark that if the largest available entanglement entropies
are not known exactly (e.g., if they are calculated from approximate
wavefunctions, or using the desnity matrix renormalization group (DMRG) technique), it might be more prudent to use a
prescription for choosing $\omega$ that actually uses more than the minimal
(three) numbers that we have used here.  In the cases reported here and in our
earlier work, \cite{our-prl-07} we have entanglement entropies calculated from
numerically exact wavefunctions, so we have no precision issues to worry about
when choosing the number of $S_{l_A}(N)$ values to use for the $N\rightarrow \infty$
extrapolation.

Once the extrapolation value is (uniquely) determined by the algorithm
outlined above, we need an estimate of the uncertainty.  We can use the bounds
($S_0$,$S_1$) derived earlier to obtain a conservative error estimate:
\[
\min(|S_1-S_{\rm BST}|,|S_{\rm BST}-S_0|)   \; ,
\]
where $S_{\rm BST}$ is the extrapolated value obtained by the BST algorithm.

\subsection{Numerical results}   \label{subsec_numerical-gamma}

\begin{figure}
\centering
 \includegraphics*[width=0.98\columnwidth]{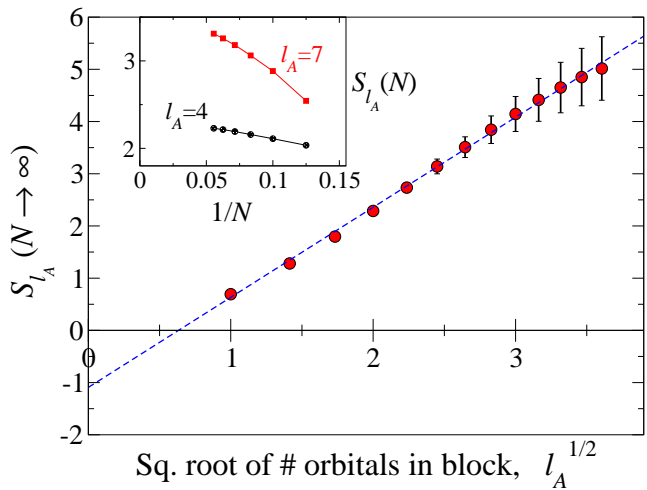}
\caption{  \label{fig_extrapolatn}
(Color online.)  Entanglement entropies in Moore-Read state wavefunctions,
  extrapolated to the thermodynamic limit.  Dashed line is a fit to
  $-\gamma+c_1\sqrt{l_A}$, with some points dropped.
%  The vertical intercept of this curve gives an estimate for $\gamma$.
%
  Inset plots $S_{l_A}$ against $1/N$ for various fixed $l_A$.
%
%  The conservative bounds ($S_0$ and $S_1$) the BST estimate for
%  $S_{l_A}(N\rightarrow\infty)$ and are shown.
%
}
\end{figure}

\emph{Moore-Read state}.
Fig.~\ref{fig_extrapolatn} shows results of numerical calculations for
the $\nu=1/2$ Moore-Read state. We used exact wavefunctions up to $N=18$
particles.
These wavefunctions were obtained by diagonalizing $\hat{L}^2$
in an $L_z=0$ Hilbert space spanned by the ``squeezed states''
\cite{Hal2}.
After numerically obtaining the entanglement entropies $S_{l_A}(N)$
from these wavefunctions, we obtain estimates and uncertainties for the
$N\rightarrow\infty$ extrapolations by the procedure outlined in the previous
subsection. The resulting data are plotted in Fig.~\ref{fig_extrapolatn}.

The linear $S_{l_A}$ versus $\sqrt{l_A}$ behavior is expected only for
large $l_A$; however our large-$l_A$ points have the greatest uncertainty.
For estimating the topological entropy, we therefore make linear fits after
discarding 0 to 5 of the smallest-$l_A$ points and/or 0 to 2 of the
largest-$l_A$ points. This results in estimates of $\gamma$ (magnitude of the
vertical intercept) scattered between 0.85 and 1.35. The error propagated into
our $\gamma$ estimate from our extrapolation uncertainties is $\sim0.3$,
larger than that obtained from this scatter.  With all this we arrive at
the result $\gamma \simeq 1.1\pm0.3$, quite consistent with the expected
value of $\gamma\simeq1.04$.

\emph{Laughlin state}.  We used the well-defined procedure of the previous
subsection to revisit our previous estimate of the topological entropy for
the $\nu=1/3$ Laughlin state.\cite{our-prl-07} To get the extrapolated
$S_{l_A}$, we now use the BST estimates rather than doing several polynomial
fits.  Dropping 0 to 4 of the smallest-$l_A$ points and 0 to 2 of the
lrgest-$l_A$ points leads to $\gamma{\simeq}0.51{\pm}0.14$, consistent with
the previously reported estimate ($0.60\pm0.15$) and with the expected value
$\gamma\approx0.55$.  The error estimate reported in
Ref.~\onlinecite{our-prl-07} only took into account this variation, due to
dropping various number of points.  There is also some error propagated from
the extrapolation uncertainty.  Using the conservative uncertainty estimate
proposed in the previous subsection gives us a more conservative and more
rigorous error estimate, $\gamma\simeq0.51\pm0.25$.

\subsection{Eigenvalue distribution for reduced density matrix}
\label{subsec_orbital-RDM-spectrum}

\begin{figure}
\centering
 \includegraphics*[width=0.98\columnwidth]{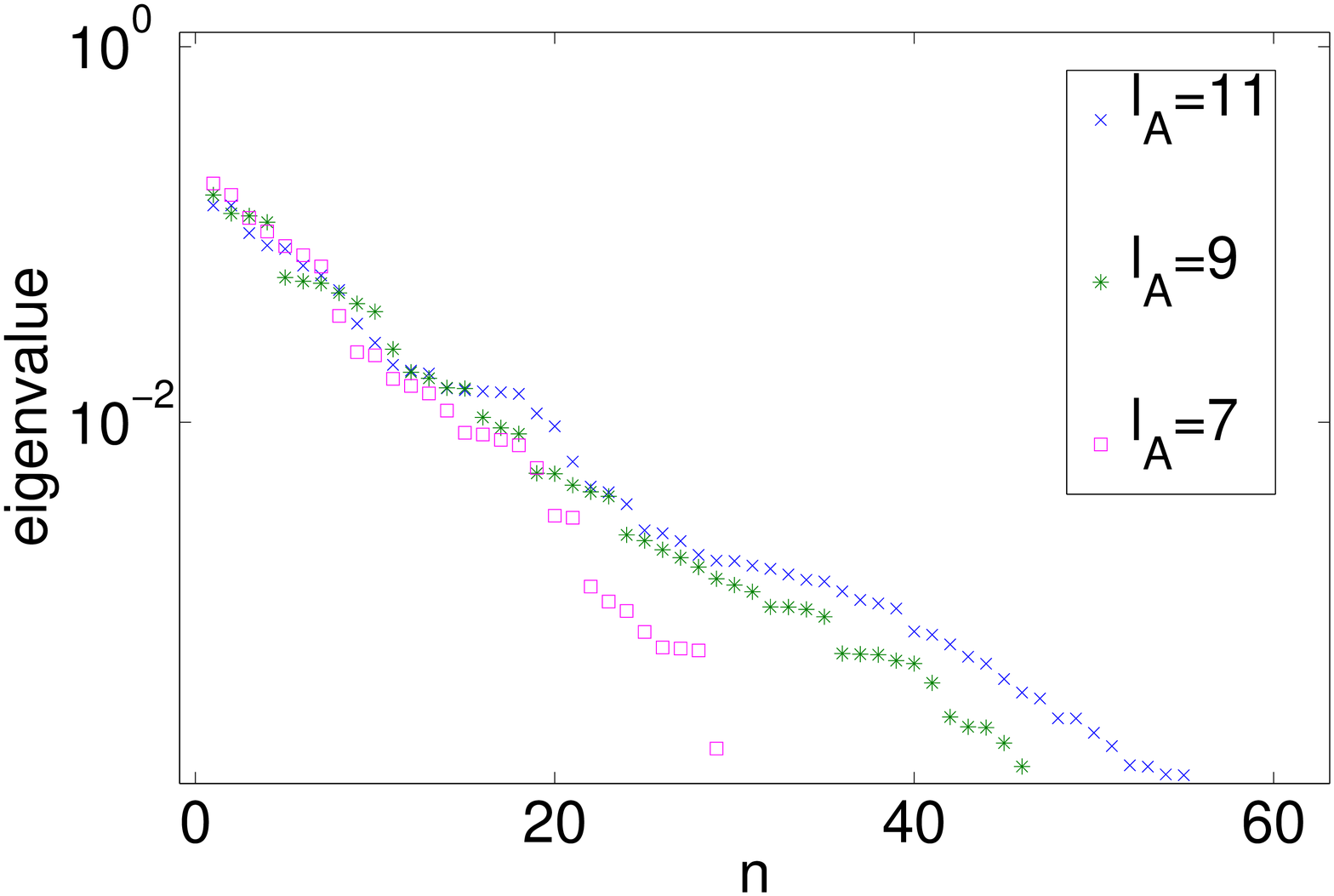}
\caption{  \label{fig_orbeigenvalues}
(Color online.) Density matrix eigenvalues in decreasing order for orbital
partitioning with $l_A$ orbitals in a block for $N=9$, $m=3$ Laughlin state.}
\end{figure}

In Fig.~\ref{fig_orbeigenvalues}, we show the largest eigenvalues of reduced
density matrices obtained by orbital or spatial partitioning.  The eigenvalues
are ordered according to decreasing magnitude and plotted on a log scale; the
resulting curves are roughly linear, suggesting a roughly exponential decay of
the eigenvalue distribution function.

It is interesting to note the complete dissimilarity of this eigenvalue
spectrum compared to the particle partitioning case discussed earlier, e.g.,
Fig.~\ref{fig_2-pcle_eigen}.  It would also be interesting to put our
observations in the context of the spectra of reduced density matrices of
many-body systems in general.  Reduced density matrices for spatially
connected blocks have been studied previously in the context of the
convergence of the DMRG algorithm; an overview is available in section III-B
of Schollw\"ock's DMRG review.  \cite{Schollwoeck_DMRG-review-05} From our
numeric data, It is difficult to say whether or not the decay of the
eigenvalue distributions is slower than exponential.

\section{Concluding remarks}  \label{sec_conclude}

We have presented a detailed study of the entanglement entropy in
abelian and non-abelian quantum Hall states, taking a paradigmatic
example of each, the $\nu=1/3$ Laughlin state and the $\nu=1/2$
Moore-Read state.

For entanglement between subsets of particles, we have demonstrated
the effects of particle-particle correlations in the deviation of the
entanglement entropies from an upper bound $S_A^F$ set by fermionic
statistics only. We presented a close upper bounds for particle
entanglement entropies for both bosonic and fermionic states. For a
particular case $n_A = N/2$, i.e. entanglement of the one half of the
system with the other, the presented upper bound is notably lower than
a similar estimate in Ref.~\onlinecite{EntngQHE_Lattore}. We showed that
the distribution of the eigenvalue spectrum of the two-particle density
matrix can be directly related to a two-particle correlation function
$g_2(r)$. This suggests that the eigenvalue distribution of
$n_A$-particle reduced density matrices can be related to $n_A$-particle
correlation functions.

For  entanglement between spatial regions, indexed by orbitals on a
spherical geometry, we have provided a thorough discussion of our
efforts to optimize the procedure for extracting the so-called
``topological entropy'' from finite-size data.  Using this procedure,
we calculated the topological entropy for the Moore-Read state, and
found our numerical result to be consistent with the expected value
$\ln\sqrt{8}$.

Our results on Laughlin and Moore-Read states indicate that the
computation of topological entanglement entropies can be used
for diagnosing topological order in FQH states that are only
available for a limited number of particles. Clearly, the practical use
of this method hinges on the accuracy by which a value for $\gamma$
can be extracted. In this context, it could be particularly interesting
to use entropies $S_{l_A}[N]$ that are generated in DMRG studies of FQH
states. 

It is interesting to note the differences between the particle and
spatial partitioning cases.  In the first case, correlation effects
decrease the particle entanglement entropy $S_{n_A}$. In the latter
case, correlation effects increase the spatial entanglement entropy;
for example, a Wigner crystal would have very high entanglement
entropy between two spatial partitions.  These differences are
particularly dramatically manifested in the difference between the
spectral structure of the reduced density matrices for the two cases.
In the particle entanglement case, the eigenvalues tend to be of the
same order, with relatively small fluctuations around some average
value. In the spatial case, the eigenvalue distribution function has
an exponential-like form, and only a small fraction of the eigenvalues
contribute to the entanglement entropy.

\begin{acknowledgments}

We thank U.~Schollw\"ock for drawing our attention to the
Bulirsh \& Stoer (BST) algorithm for extrapolations
and we acknowledge discussions with E.~Ardonne (on eq.~(\ref{scriptDAS})),
P.~Calabrese. We thank  A.~Feiguin for discussions and providing us with DMRG
data for large-size Laughlin wave functions. The research of OZ and KS
is financially supported by the Stichting voor Fundamenteel Onderzoek
der Materie (FOM) of the Netherlands. EHR is supported by US DOE under
contract DE-FG03-02ER-45981. KS and EHR acknowledge the Institute
for Pure and Applied Mathematics (IPAM) at Los Angeles, where part of
this work was done.

\end{acknowledgments}

\appendix

\section{2-particle reduced density matrix and correlation functions}
\label{eigenvalues_correlations}

In this appendix we derive  Eq.~(\ref{rho2_theta}).
The two-particle reduced density matrix may be written as
\begin{equation}
\rho_2 = \sum_{l,m} \lambda_{lm} | l,m \rangle \langle l,m | \; .
\end{equation}
where $\ket{l,m}$ is a state of two particles with total angular momentum $l$
and projection $m$.  The rotational symmetry of the sphere ($\rho_2$ must be
invariant under rotations) indicates that the eigenvalues $\lambda_{lm}$ do
not depend on the projection $m$, i.e, are grouped in $SU(2)$ multiplets;
$\lambda_{lm}=\lambda_{l}$.

To get the correlation function, we need to compute the density matrix in a
basis of polar spherical coordinates,
\begin{equation}
 \rho_2(\theta_1, \theta_2, \phi_1 , \phi_2)
  = \bra{\theta_1, \theta_2, \phi_1 , \phi_2} \rho_2
    \ket{\theta_1, \theta_2, \phi_1 , \phi_2} \; ,
\end{equation}
where $\theta_i$ and $\phi_i$ are coordinates of particles on a sphere.
Rotational symmetry requires that $\rho_2$ depends only on the
angular distance $\theta$ between particles:
\begin{equation}
\cos \theta = \cos(\phi_1 - \phi_2) \sin \theta_1 \sin \theta_2
              + \cos \theta_1 \cos \theta_2 \ .
\end{equation}

Now the unnormalized wavefunction of two particles with total angular
momentum $l$ can be written as \cite{Haldane_FQHE_PRL83}
\begin{equation}
\Psi_l = \prod_{i=1,2}(\alpha^* u_i + \beta^* v_i)^{l}
                      (u_1 v_2 -u_2 v_1)^{2L-l} \; ,
\end{equation}
where $L$ is the angular momentum of each particle, $u_i = \cos
\frac{\theta_i}{2} e^{i \phi_i/2}$, $v_i = \sin \frac{\theta_i}{2} e^{- i
\phi_i/2}$, and $\alpha$, $\beta$ fix the center of mass of the two
particles.  The same wavefunction can also be written as
\begin{equation}
\Psi_l = \bra{\theta_1, \theta_2, \phi_1, \phi_2} \hat{D}(\alpha, \beta)
         \ket{l,l} \; ,
\end{equation}
where $\hat{D}(\alpha, \beta)$ is a rotation operator. Let us write
$\ket{\Omega} = \hat{D}(\alpha, \beta) \ket{l,l}$, which is a familiar
coherent state.
Using the identity
\begin{equation}
\frac{2l + 1}{4 \pi} \int d \Omega \ket{\Omega} \bra{\Omega}
= \sum_m \ket{l,m} \bra{l,m}
\end{equation}
one obtains
\begin{multline}
\sum_{m} |\langle\theta_1, \theta_2, \phi_1, \phi_2 \ket{l,m}|^2 = \\
 \frac{2l + 1}{4 \pi} \int d \Omega~
 |\langle \theta_1, \theta_2, \phi_1, \phi_2 \ket{\Omega}|^2 \; .
\end{multline}
%
%Because of the rotational symmetry,
The normalization of
$\langle \theta_1, \theta_2, \phi_1, \phi_2 \ket{\Omega}$
is equal to %that of
%$\langle \theta_1, \theta_2, \phi_1, \phi_2 \ket{l,l}$, which is
%given by
\cite{Fano}
\begin{equation}
N_l =% \int_0^{\pi} d\theta_1 \int_0^{\pi} d \theta_2
     % \int_0^{2\pi} d \phi_1 \int_0^{2\pi}  d\phi_2
%\\[2mm] \quad
%\sin \theta_1 \sin \theta_2 |u_1|^{2l} |u_2|^{2l} |u_1 v_2 -u_2 v_1|^{4L-2l}
%\\[2mm]
 \frac{16 \pi^2 (l!)^2}{(2l+1)!}\frac{(2L-l)!(2L+l+1)!}{(2L+1)!^2} \; .
\end{equation}

We need to compute
\begin{multline}
 I_l \equiv \int d \Omega~ |\langle \theta_1, \theta_2, \phi_1, \phi_2
 \ket{\Omega}|^2 \\
  = \int d \Omega_{\alpha, \beta}
  \prod_{i=1,2}|\alpha^* u_i + \beta^* v_i|^{2l}
               |u_1 v_2 -u_2 v_1|^{4L-2l} \; .
\end{multline}
Defining $ a \equiv |u_1 v_2 -u_2 v_1|^2 = (1 - \cos \theta)/2$
and $b \equiv |u_1^*u_2+v_2 v_1^{*}|^2 = (1 + \cos \theta)/2$, one finds
\begin{multline}
I_l(\theta) = \frac{4 \pi (l!)^2}{(2l+1)!} a^{2L-l} \sum_{k=0}^{l}
\frac{(l+k)!}{(l-k)!(k!)^2} a^{l-k} b^k \\
=  \frac{4 \pi (l!)^2}{(2l+1)!} a^{2L}
\! \! \phantom{x}_2 F_1 (-l,l+1,1,-b/a)
\; .
\end{multline}
%
%% %
%% Using the identity
%% %
%% \begin{equation}
%% \sum_{k=0}^{l} \frac{(l+k)!}{(l-k)!(k!)^2} x^k = \! \! \phantom{x}_2 F_1(-l,l+1,1 -x) \;,
%% \end{equation}
%% %
%% it can be rewritten as
%% \begin{equation}
%% I_l(\theta) =  \frac{4 \pi (l!)^2}{(2l+1)!} a^{2L}
%% \! \! \phantom{x}_2 F_1 (-l,l+1,1,-b/a) \;.
%% \end{equation}
%% %
%
One can see that $I_l$ is normalized as
\begin{equation}
\int_0^{\pi} d \theta \sin \theta I_l(\theta) = \frac{N_l}{2 \pi} \; .
\end{equation}
Introducing the function
\begin{multline}
R_l(\theta) \equiv \frac{I_l(\theta)}{N_l} =\\
\frac{(2L+1)!^2}{4 \pi (2L-l)! (2L+l+1)!} a^{2L} \! \! \phantom{x}_2
F_1(-l,l+1,1,-b/a)
\label{Rtheta}
\end{multline}
we finally write the 2-body correlator as
\begin{equation}
\rho_2(\theta) = \sum_{l} \lambda_l \frac{2l+1}{4\pi} R_l(\theta) \ .
\end{equation}

\section{Entanglement entropies; numerical results with orbital partitioning}

\begingroup
%\squeezetable
\begin{table*}
\begin{tabular}{|c||c|c|c|c|c|c|c|c|}
\hline
$l_A$ &
$N= 4$&
$N= 6$&
$N= 8$&
$N= 10$&
$N= 12$&
$N= 14$&
$N= 16$&
$N= 18$
\\
\hline
\hline
1&
0.636514&
0.673012&
0.682908&
0.686962&
0.689009&
0.690186&
0.690923&
0.691416
\\
\hline
2&
1.09861&
1.15777&
1.23519&
1.24961&
1.25796&
1.2663&
1.27095&
1.27385
\\
\hline
3&
{\bf 1.09861}&
1.49971&
1.65513&
1.7002&
1.72587&
1.74591&
1.75859&
1.76716
\\
\hline
4&
&
1.76712&
2.0355&
2.11077&
2.15836&
2.19257&
2.21444&
2.22959
\\
\hline
5&
&
{\bf 1.88152}&
2.31807&
2.44895&
2.52664&
2.58123&
2.61639&
2.64089
\\
\hline
6&
&
&
2.48295&
2.70298&
2.82429&
2.90754&
2.96113&
2.99855
\\
\hline
7&
&
&
{\bf 2.54282}&
2.88348&
3.06178&
3.18053&
3.257&
3.31045
\\
\hline
8&
&
&
&
2.98996&
3.24156&
3.40254&
3.50653&
3.57927
\\
\hline
9&
&
&
&
{\bf 3.02558}&
3.36728&
3.57792&
3.71402&
3.80922
\\
\hline
10&
&
&
&
&
3.44165&
3.71089&
3.88358&
4.00428
\\
\hline
11&
&
&
&
&
{\bf 3.46601}&
3.80408&
4.01821&
4.16743
\\
\hline
12&
&
&
&
&
&
3.85949&
4.12054&
4.30135
\\
\hline
13&
&
&
&
&
&
{\bf 3.87774}&
4.19219&
4.408
\\
\hline
\end{tabular}
\caption{\label{table_MR_entropies}
Orbital-partitioning entanglement entropies $S_{l_A}[N]$ for
the $\nu=1/2$ Moore-Read state.}
\end{table*}
\endgroup

\begingroup
%\squeezetable
\begin{table*}
\begin{tabular}{|c||c|c|c|c|c|c|c|c|}
\hline
$l_A$  & $N=3$ & $N=4$ & $N=5$ & $N=6$ & $N=7$ & $N=8$ & $N=9$ & $N=10$ \\
\hline\hline
1 & 0.682908 & 0.673012 & 0.666278 & 0.661563 & 0.65811 & 0.655482 &
0.653418 & 0.651757 \\
\hline
2 & 1.00424 & 1.05492 & 1.07339 & 1.0822 & 1.08707 & 1.09006 & 1.09202
&  1.09337  \\
\hline
3 & {\bf 1.27703} & 1.3944 & 1.44547 & 1.46998 & 1.48491 & 1.49466 & 1.50147
&  1.50648  \\
\hline
4 & {\bf 1.27703} & 1.59387 & 1.70596 & 1.76262 & 1.79672 & 1.81933 &
1.8353 &  1.84717  \\
\hline
5 &  & {\bf 1.66184}  & 1.87354 & 1.97796 & 2.03976 & 2.08068
& 2.10962 &  2.13113  \\
\hline
6 &  &  & {\bf 1.9523}  & 2.1202 & 2.21872 & 2.28347 &
2.32919 &  2.36314  \\
\hline
7 &   &  & {\bf 1.9523}  & 2.20091 & 2.3448 & 2.43838 & 2.50425 &
2.55307  \\
\hline
8 &   &  &  & {\bf 2.22768} & 2.42555 & 2.55286 & 2.64196 &
2.70783 \\
\hline
9 &   &  &  &  & {\bf 2.46451}  & 2.63139 & 2.74713
&  2.83233  \\
\hline
10 &   &   &  &  & {\bf 2.46451} & 2.67761 & 2.82363 &
2.93049 \\
\hline
11 &   &   &  &  &  & {\bf 2.69293} &  2.87356  &  3.00467
 \\
\hline
12 &   &   &  &  &  &  & {\bf 2.89817} &  3.05653 \\
\hline
\end{tabular}
\caption{\label{table_Laughlin_entropies}
Orbital-partitioning entanglement entropies $S_{l_A}[N]$ for the
$\nu=1/3$ Laughlin state.}

\end{table*}
\endgroup

In tables \ref{table_MR_entropies} and \ref{table_Laughlin_entropies} we list
orbital-partitioning entanglement entropies calculated using the numerical
wavefunctions.  For each wavefunction (each colmun), the entanglement
entropies are only listed up to their maximum value, typeset in bold, because
the values after this are determined by the symmetry
$S_{l_A}=S_{N_{\phi}+1-l_A}$.  The extrapolation procedure of
Sec.~\ref{subsec_extrapoln-issues} involves extrapolating each row of numbers
to $N\rightarrow\infty$.

Note that, in each column (for a particular $N$), the omitted part after the
midpoint (in bold) is a \emph{decreasing} function of $l_A$ and thus does not
give useful information about the thermodynamic limit of $S_{l_A}$.  In the
extrapolation, it is therefore important to avoid values from these parts of
the table.  We therefore restrict ourselves to values which in tables
\ref{table_MR_entropies} and \ref{table_Laughlin_entropies} are to the right
of (i.e., above) the diagonal line through the midpoint numbers typeset in
bold.

%%%%%%%%%%%%%%%%%%%%%%%%%%%%%%%%%%%%%%%%%%%%%%%%%%%%%%%%%%%%%%%%%%


\begin{thebibliography}{99}

\bibitem{WenNiu_PRB}
X.~G.~Wen and Q.~Niu, Phys.~Rev.~B {\bf 41}, 9377 (1990);
X.~G.~Wen, Phys.~Rev.~B {\bf 41}, 12838 (1990);
X.~G.~Wen, Phys.~Rev.~B {\bf 44}, 2664 (1991).

\bibitem{Moore-Read_NP91}
G.~Moore and N.~Read, Nucl.~Phys.~B{\bf 360}, 362 (1991).

\bibitem{Read-Rezayi_PRB99}
N.~Read and E.~Rezayi, Phys.~Rev.~B {\bf 59}, 8084 (1999).

\bibitem{Ardonne_Schoutens99}
E.~Ardonne and K.~Schoutens, Phys.~Rev.~Lett. {\bf 82}, 5096 (1999).

\bibitem{topological-quantum-computing}
A. Yu. Kitaev, Annals Phys. {\bf 303}, 2 (2003); M. Freedman, M.
Larsen, and Z. Wang, Commun. Math. Phys. 227, 605 (2002); N. E.
Bonesteel, L. Hormozi, G. Zikos, S.H. Simon, Phys. Rev. Lett. {\bf
95}, 140503 (2005); S. Das Sarma, M. Freedman, C. Nayak, Phys.
Rev. Lett. {\bf 94}, 166802 (2005).

\bibitem{VidalLatorreRicoKitaev_PRL03}
G.~Vidal, J.~I.~Latorre, E.~Rico and A.~Kitaev,
Phys.~Rev.~Lett.~{\bf 90}, 227902 (2003).

\bibitem{Entng-in-spin-chains}
J.~I.~Latorre, E.~Rico, and G.~Vidal,
Quantum Information and Computation, {\bf 4}, 48 (2004);
V.E.~Korepin,  Phys. Rev. Lett. {\bf 92}, 096402 (2004);
V.~Popkov and M.~Salerno,  Phys. Rev. A {\bf 71}, 012301 (2005).

\bibitem{Cardy_JStatMech04}
P.~Calabrese and J.~Cardy, J. Stat. Mech. {\bf 0406}, 002 (2004).

\bibitem{Eisert_2002}
K.~Audenaert, J.~Eisert, M.~B.~Plenio, R.~F.~Werner, Phys. Rev. A {\bf 66},
042327 (2002).

\bibitem{our-prl-07}
M.~Haque, O.~Zozulya, and K.~Schoutens,
Phys. Rev. Lett.~{\bf 98}, 060401 (2007).

\bibitem{Preskill-Kitaev_PRL06}
A.~Kitaev and J.~Preskill,
Phys.~Rev.~Lett.~{\bf 96}, 110404 (2006).

\bibitem{Levin-Wen_PRL06}
M.~Levin and X.~G.~Wen,
Phys.~Rev.~Lett.~{\bf 96}, 110405 (2006).

\bibitem{Read-Rezayi96}
N.~Read and E.~Rezayi, Phys.~Rev.~B {\bf 54}, 16864 (1996).

\bibitem{Read_2000-overview}
N.~Read, Physica {\bf 298B}, 121 (2001).

\bibitem{HolzheyWilczek_NPB95}
C.~Holzhey, F.~Larsen, and F.~Wilczek,
Nucl.~Phys.~B {\bf 424}, 443 (1995).

\bibitem{Tsinghua_FQHE_PRA02}
B.~Zeng, H.~Zhai, and Z.~Xu, Phys.~Rev.~A {\bf 66}, 042324 (2002).

\bibitem{EntngQHE_Lattore}
S.~Iblisdir, J.~I.~Latorre, and R.~Or\'us,
Phys. Rev. Lett. {\bf 98}, 060402 (2007).

\bibitem{Pcle_entanglement_Santachiara}
R.~Santachiara, F.~Stauffer, D.~Cabra, cond-mat/0610402;
P.~Calabrese and M.~Mintchev, cond-mat/0703117.

\bibitem{Misguich_dimer-gamma_dec06}
S.~Furukawa and G.~Misguich, cond-mat/0612227.

\bibitem{Haldane_FQHE_PRL83}
F.~D.~M.~Haldane, Phys.~Rev.~Lett.~{\bf 51}, 605 (1983).

\bibitem{ArovasAuerbachHaldane_PRL88}
D.~P.~Arovas, A.~Auerbach, and F.~D.~M.~Haldane,
Phys.~Rev.~Lett.~{\bf 60}, 531 (1988).

\bibitem{NAcounting}
V.~Gurarie and E.~Rezayi, Phys.~Rev.~B {61}, 5473 (2000);
E.~Ardonne, N.~Read, E.~Rezayi and K.Schoutens, Nucl.~Phys.~B
{\bf 607}, 549 (2001); E.~Ardonne, J.~Phys.~A {\bf 35}, 447 (2002).

\bibitem{Srednicki_PRL93}
M.~Srednicki, Phys.~Rev.~Lett. {\bf 71}, 666 (1993);
M.~M.~Wolf, F.~Verstraete, M.~B.~Hastings, J.~I.~Cirac, arXiv: 0704.3906;
M.~B.~Plenio, J.~Eisert, J.~Drei\ss ig, M.~Cramer, Phys.~Rev.~Lett. {\bf 94}, 060503 (2005).

\bibitem{FendleyFisherNayak_JStatPhys07}
P.~Fendley, M.~P.~A.~Fisher, and C.~Nayak,
J.~Stat. Phys. {\bf 126}, 1111 (2007).

\bibitem{xtrap_BulirshStoer64}
R.~Bulirsh and J.~Stoer, Numer.~Math.~{\bf 6}, 413 (1964).

\bibitem{xtrap_HenkelSchuetz88}
M.~Henkel and G.~Sch\"utz, J.~Phys.~A {\bf 21}, 2617 (1988).

\bibitem{Hal2}
F.D.M.~Haldane, unpublished.

\bibitem{Schollwoeck_DMRG-review-05}
U.~Schollw\"ock, Rev.~Mod.~Phys. {\bf  77}, 259 (2005).



\bibitem{Fano}
G.~Fano, F.~Ortolani, E.~Colombo, Phys.~Rev.~B {\bf 34}, 2670, (1986).

\end{thebibliography}
\end{document}